\documentclass[twocolumn,prl,showpacs,floatfix,altaffilletter]{revtex4}
\usepackage{graphicx}
\usepackage{bm}
\usepackage{amsmath}
\usepackage{amsfonts}
\usepackage{amssymb}
\begin{document}
\title{\textbf{Spin waves and revised crystal structure of honeycomb iridate Na$_{2}$IrO$_{3}$}}
\author{S.K. Choi$^{1}$, R. Coldea$^1$, A.N. Kolmogorov$^{2}$, T.
Lancaster$^{1a}$, I.I. Mazin$^{3}$, S.J. Blundell$^{1}$, P.G.
Radaelli$^1$, Yogesh Singh$^{4b}$, P. Gegenwart$^4$, K.R.
Choi$^5$, S.-W. Cheong$^{5,6}$, P.J. Baker$^7$, C. Stock$^7$, J.
Taylor$^7$} \affiliation{$^{1}$Clarendon Laboratory, University of
Oxford, Parks Road, Oxford OX1 3PU, U.K.}
\affiliation{$^{2}$Department of Materials, University of Oxford,
Parks Road, Oxford OX1 3PH, U.K.} \affiliation{$^{3}$Code 6393,
Naval Research Laboratory, Washington, DC 20375, USA}
\affiliation{$^{4}$I. Physikalisches Institut,
Georg-August-Universit\"{a}t G\"{o}ttingen, D-37077 G\"{o}ttingen,
Germany} \affiliation{$^{5}$l\_PEM, Pohang University of Science
and Technology, Pohang 790-784, Korea} \affiliation{$^{6}$R-CEM
and Department of Physics and Astronomy, Rutgers University,
Piscataway, New Jersey 08854, USA} \affiliation{$^{7}$ISIS
Facility, Didcot, Oxfordshire, OX11 0QX, U.K.}
\date{\today}


\begin{abstract}
We report inelastic neutron scattering measurements on
Na$_{2}$IrO$_{3}$, a candidate for the Kitaev spin model on the
honeycomb lattice. We observe spin-wave excitations below 5 meV
with a dispersion that can be accounted for by including
substantial further-neighbor exchanges that stabilize zig-zag
magnetic order. The onset of long-range magnetic order below
$T_{\mathrm{N}}=15.3$~K is confirmed via the observation of
oscillations in zero-field muon-spin rotation experiments.
Combining single-crystal diffraction and density functional
calculations we propose a revised crystal structure model with
significant departures from the ideal 90$^{\circ}$ Ir-O-Ir bonds
required for dominant Kitaev exchange.
\end{abstract}
\maketitle

Transition metal oxides of the $5d$ group have recently attracted
attention as candidates to exhibit novel electronic ground states
stabilized by the strong spin-orbit (SO) coupling, including
topological band- or Mott-insulators \cite{balents}, quantum spin
liquids \cite{jackeli}, field-induced topological order
\cite{trebst}, topological superconductors \cite{Ashvin} and
spin-orbital Mott insulators \cite{Sr2IrO4}. The compounds ${\cal
A}$$_{2}$IrO$_{3}$ (${\cal A}$=Li, Na) \cite{gegenwart,Yogesh}, in
which edge-sharing IrO$_{6}$ octahedra form a honeycomb lattice
[see Fig.~\ref{fig1}b)], have been predicted to display novel
magnetic states for composite spin-orbital moments coupled via
frustrated exchanges. The exchange between neighboring Ir moments
(called $\bm{S}_{i,j}$, $S$=1/2) is proposed to be \cite{jackeli}
\begin{equation}
\mathcal{H}_{ij}=-J_{\mathrm{K}}S_{i}^{\gamma}S_{j}^{\gamma}+J_1\bm{S}_{i}\cdot\bm{S}_{j},\label{eq_HK}
\end{equation}
where $J_{\mathrm{K}}>0$ is an Ising ferromagnetic (FM) term
arising from superexchange via the Ir-O-Ir bond, and $J_1>0$ is
the antiferromagnetic (AFM) Heisenberg exchange via direct Ir-Ir
5$d$ overlap. Due to the strong spin-orbital admixture the Kitaev
term $J_{\mathrm{K}}$ couples only the components in the direction
$\gamma$, normal to the plane of the Ir-O-Ir bond
\cite{JK,jackeli-bonds}. Because of the orthogonal geometry,
different spin components along the cubic axes ($\gamma=x,y,z$) of
the IrO$_{6}$ octahedron are coupled for the three bonds emerging
out of each site in the honeycomb lattice. This leads to the
strongly-frustrated Kitaev-Heisenberg (KH) model \cite{jackeli},
which has conventional N\'{e}el order [see Fig.~\ref{ins}a)] for
large $J_1$, a stripy collinear AFM phase [see Fig.~\ref{ins}c)]
for $0.4 \lesssim \alpha \lesssim 0.8$, where
$\alpha=J_{\mathrm{K}}/\left(J_{\mathrm{K}}+2J_1\right)$ (exact
ground state at $\alpha=1/2$), and a quantum spin liquid with
Majorana fermion excitations \cite{kitaev} at large
$J_{\mathrm{K}}$ ($\alpha \gtrsim 0.8$). In spite of many
theoretical studies
\cite{trebst,jackeli,shitade,Jin,YBK,Kimchi,Ashvin} very few
experimental results are available for ${\cal A}_2$IrO$_3$
\cite{gegenwart,Yogesh,hill}. Evidence of unconventional magnetic
order in Na$_2$IrO$_3$ came from resonant xray scattering
\cite{hill} which showed magnetic Bragg peaks at wavevectors
consistent with either an in-plane zig-zag or stripy order [see
Figs.~\ref{ins}b-c)].

Measurements of the spin excitations are very important to
determine the overall energy scale and the relevant magnetic
interactions, however because Ir is a strong neutron absorber
inelastic neutron scattering (INS) experiments are very
challenging. Using an optimized setup we here report the first
observation of dispersive spin wave excitations of Ir moments via
INS. We show that the dispersion can be quantitatively accounted
for by including substantial further-neighbor in-plane exchanges,
which in turn stabilize zig-zag order. To inform future {\em ab
initio} studies of microscopic models of the interactions we
combine single-crystal xray diffraction with density functional
calculations to determine precisely the oxygen positions, which
are key in mediating the exchange and controlling the spin-orbital
admixture via crystal field effects. We propose a revised crystal
structure with much more symmetric IrO$_6$ octahedra, but with
substantial departures from the ideal 90$^{\circ}$ Ir-O-Ir bonds
required for dominant Kitaev exchange \cite{jackeli-bonds}, and
with frequent structural stacking faults. This differs from the
currently-adopted model, used by several band-structure
calculations \cite{hill,YBK}, with asymmetrically-distorted
IrO$_6$ octahedra, with Ir-O bonds differing in length by more
than 20\%, improbably large in the absence of any Jahn-Teller
interaction, and with the shortest Ir-O bond length below 2~\AA,
highly unlikely for a large ion like Ir$^{4+}$. We show that the
previously proposed structure is unstable with large unbalanced
ionic forces, and when allowed to relax it converges to a
higher-symmetry structure.

\begin{figure}[t!]
\begin{center}
\includegraphics[width=8.5cm]{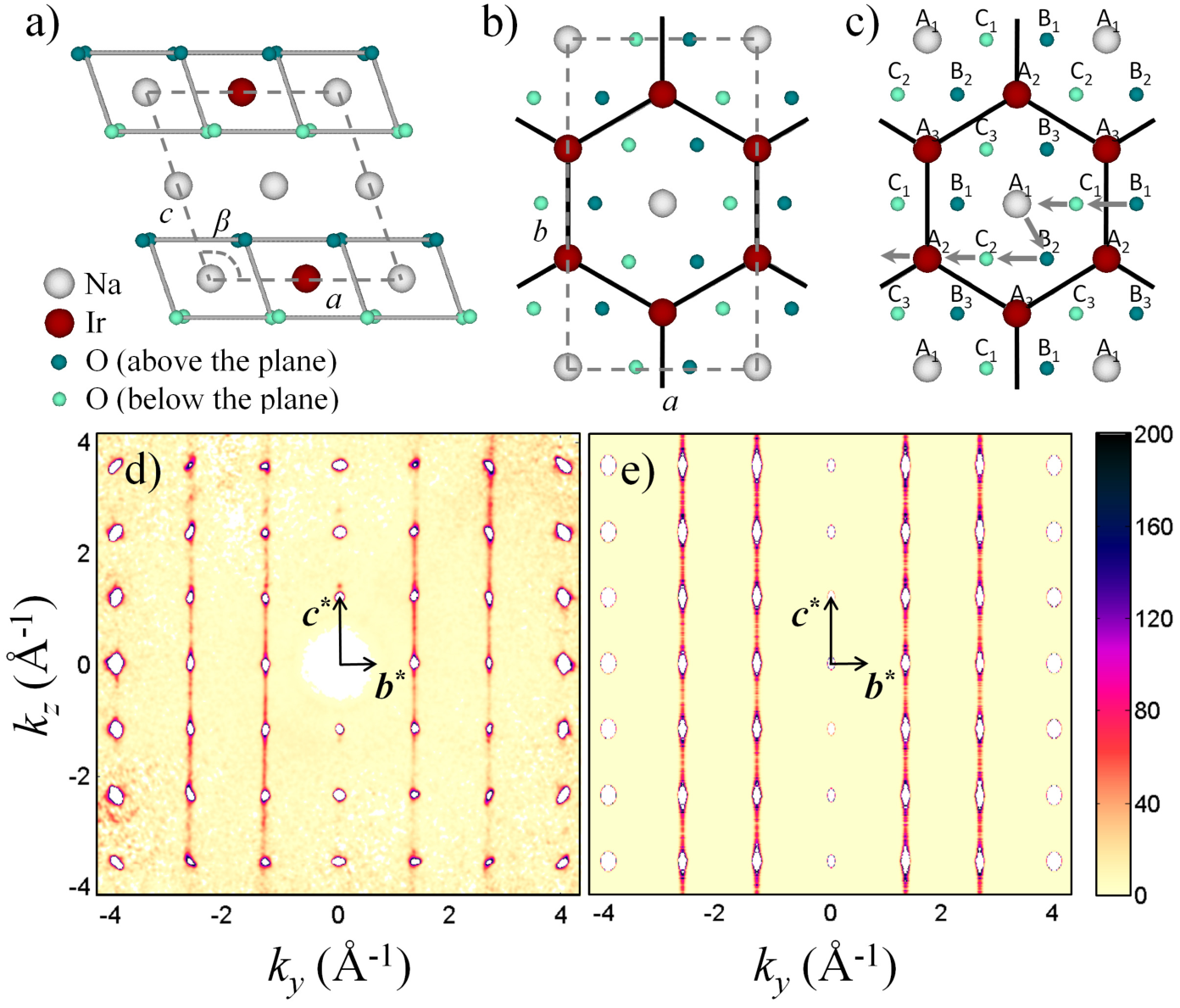}
\end{center} \caption{(Color
online) a) Layer stacking along the monoclinic $c$-axis with an
in-plane offset along $a$ (dashed box is the C$2/m$ unit cell). b)
Basal layer ($z=0$) showing the Ir honeycomb lattice.
c) Diagram to illustrate the layer stacking in the ideal honeycomb
lattice. Ideal stacking of layers and stacking faults are
explained in the text. d) Xray diffraction intensity in the
($0,k,l$) plane showing rods of diffuse scattering in between
structural Bragg peaks along ${\bm c}^{*}$ with selection rule
$h+k=2n$ and $k=3m+1$ or $3m+2$ ($n,m$ integers) modelled in e) by
frequent in-plane translational stacking faults of the type shown
by the thick arrows in c).} \label{fig1}
\end{figure}

As other ``213'' honeycomb oxides, Na$_2$IrO$_3$ has an
alternating stacking of hexagonal layers of edge-sharing NaO$_{6}$
octahedra and similar layers where two-thirds of Na are replaced
by Ir to form a honeycomb lattice with Na in the center [see
Fig.~\ref{fig1}b)]. To determine the precise structure xray
diffraction was performed on a flux-grown single crystal of
Na$_{2}$IrO$_{3}$ \cite{gegenwart,supplementary}. The diffraction
pattern showed sharp Bragg peaks which could be indexed by a
monoclinic unit cell [see Fig.~\ref{fig1}a)] derived from a parent
rhombohedral structure with an ideal repeat every three layers.
The monoclinic distortion leads to an in-plane shift of successive
Ir honeycombs differing by 1.2\% from the ideal value
[$-c\cos\beta$ compared to $a/3$, see Fig.~\ref{fig1}a)], well
above our instrumental resolution, which enabled us to determine
that our sample was a single monoclinic domain. The detailed
refinement \cite{supplementary} was performed using both the
published C$2/c$ (No.~15) unit cell with 15 refined atomic
positions leading to values somewhat similar to Ref.\
\cite{gegenwart}, and an alternative, higher-symmetry and half the
unit cell volume, C$2/m$ model (No.~12, shown in
Figs.~\ref{fig1}a-b) (as found for the related Li$_{2}$IrO$_{3}$
\cite{Li2IrO3}), with only 7 refined atomic positions listed in
Table\ \ref{tab_unitcell}. Other structural motifs reported for
``213'' honeycomb oxides \cite{ICSD database} including
Na$_{2}$PtO$_{3}$, Li$_{2}$TeO$_{3}$, Na$_{2} $TbO$_{3}$ were also
tried but did not provide a good fit. We also tested for Ir/Na
site admixture but this did not improve the agreement with data.

The C$2/c$ structure can be described as a ``supercell'' obtained
from the C$2/m$ structure by small displacements of atoms (of
order a few $\%$ of the unit cell dimensions) leading to a doubled
unit cell volume. Although C$2/m$ and C$2/c$ gave comparable
agreement with the main Bragg peaks, the larger C$2/c$ unit cell
should be manifested experimentally by the appearance of new
``superstructure'' peaks at positions such as
(odd,odd,half-integer) in the small unit cell description
(C$2/m$). These superlattice peaks, however, were not observed in
the data \cite{supplementary}, ruling out the C$2/c$ model.
Furthermore, in structural optimization calculations using {\small
VASP} \cite{VASP,supplementary} (also confirmed by an all-electron
LAPW code \cite{WIEN}) we find that the C$2/c$ structural model,
which has asymmetrically-distorted IrO$_{6}$ octahedra, is
unstable: (i) the forces on oxygen are very large, exceeding 3
eV/A for the published C$2/c$ cell \cite{gegenwart} and (ii) when
the structure is allowed to relax the oxygens move such as to
recover the more symmetric C$2/m$ structure with the Ir-O
distances converging to within 1.1$\%$ of the experimentally
refined values in Table\ \ref{tab_unitcell}. The IrO$_{6}$
octahedra are much more symmetric in the C$2/m$ model with Ir-O
distances and Ir-O-Ir bond angles ranging from 2.06 to 2.08~\AA,
and 98 to 99.4$^{\circ}$, respectively, compared to the wider
ranges 1.99 to 2.43~\AA~, and 91 to 98$^{\circ}$ proposed before
\cite{gegenwart}.

\begin{table}[ptb]
\caption{Structural parameters extracted from single-crystal xray
data at 300~K. (C$2/m$ space group, $a=5.427(1)$~\AA ,
$b=9.395(1)$~\AA , $c=5.614(1)$~\AA , $\beta
=109.037(18)^{\circ}$, $Z$=4). All sites are fully occupied. $U$
is the isotropic displacement. The goodness-of-fit was 2.887
($R_{int}=0.1247$, $R_{\sigma}=0.0584$) \cite{supplementary}.}
\label{tab_unitcell}
\par
\begin{center}%
\begin{tabular}
[c]{llllll}\hline\hline
Atom & Site & $x$ & $y$ & $z$ & $U$(\AA $^{2}$)\\\hline
Ir ~~ & 4$g$ ~~ & 0.5 ~~ & 0.167(1) ~~ & 0 ~~ & 0.001(1)\\
Na1 ~~ & 2$a$ ~~ & 0 ~~ & 0 ~~ & 0 ~~ & 0.001(6)\\
Na2 ~~ & 2$d$ ~~ & 0.5 ~~ & 0 ~~ & 0.5 ~~ & 0.009(7)\\
Na3 ~~ & 4$h$ ~~ & 0.5 ~~ & 0.340(2) ~~ & 0.5 ~~ & 0.009(6)\\
O1 ~~ & 8$j$ ~~ & 0.748(6) ~~ & 0.178(2) ~~ & 0.789(6) ~~ & 0.001(6)\\
O2 ~~ & 4$i$ ~~ & 0.711(7) ~~ & 0 ~~ & 0.204(7) ~~ &
0.001(7)\\\hline\hline
\end{tabular}
\end{center}
\end{table}

In addition to sharp Bragg peaks, visible diffuse ``rods'' of
scattering were also observed [see Fig.~\ref{fig1}d)] and could be
quantitatively understood [compare with calculation in
Fig.~\ref{fig1}e)] in terms of a structural model that allows for
the possibility of faults in the stacking sequence along the
$c$-axis. The stacking of atomic layers can be easily visualized
with reference to projections in the basal plane
[Fig.~\ref{fig1}c)], where A defines a nominal hexagonal lattice
(made up of three triple-cell sublattices A$_1$-A$_3$), and B and
C are also hexagonal lattices with positions in the center of a
triangles of A sites. The atomic stacking is always in the ABC
sequence to minimize the interlayer Coulomb energy, i.e.
Ir-O-Na-O-Ir-O is A$_{1}$-B-C-A-B$_{1}$-C. Only Ir layers have a
sublattice index, indicating the position of the Na at the
honeycomb center, as the other atomic layers are full hexagonal
lattices. However, if neighboring Ir layers are only weakly
interacting (as they are separated by a hexagonal NaO$_2$ layer)
then the second Ir layer could be shifted to another position on
the B-lattice, say B$_{2}$ [thick arrows in Fig.~\ref{fig1}c)] or
B$_{3}$, with only minimal energy cost, as that would not affect
the bonding with the fully hexagonal NaO$_{2}$ layers below and
above. To quantitatively verify this idea, we performed structural
optimization calculations using VASP \cite{supplementary} in an
extended unit cell to include a stacking fault of the type
illustrated in Fig.~\ref{fig1}c) and found that the energy cost of
a stacking fault is extremely small, below 0.1 meV/\AA$^2$,
explaining why such stacking faults are very likely to occur.

The calculated scattering for such a microscopic model
\cite{supplementary} indeed reproduces well the selection rule for
where diffuse scattering occurs in Fig.~\ref{fig1}d-e). In
particular there is no diffuse scattering along $(00l)$, as this
corresponds to adding all layers in phase irrespective of their
in-plane translations. Also there is no diffuse scattering along
$(0,6n,l)$ ($n$ integer), as again layers add in phase because the
two allowed in-plane translations have a phase factor equal to a
multiple of $2\pi$. We use the strength of the diffuse scattering
integrated between (020) and (021) relative to the intensity of
the (020) peak (to have similar absorption factor), obtained
experimentally as $\simeq$ 0.42, to estimate the probability for
stacking faults $p\simeq9\%$, this means that on average one fault
occurs every $1/p\simeq10$ layers. We measured over 30 crystals
from a batch and all showed diffuse scattering, suggesting that
this is a common structural feature.

\begin{figure}[tbp]
\begin{center}
\includegraphics[width=8.5cm]{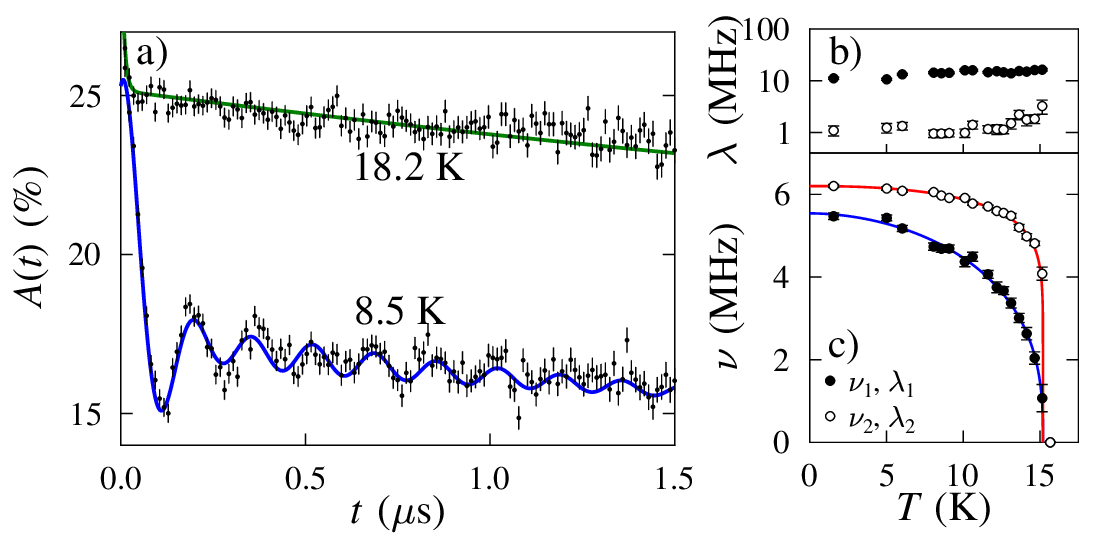}
\end{center}
\caption{(Color online) a) ZF $\mu^{+}$SR spectra on a
polycrystalline sample of Na$_{2}$IrO$_{3}$ above and below
$T_{N}$. Solid lines are (top) guide to the eye and (bottom) a fit
described in the text. b,c) Fitted parameters as a function of
temperature.} \label{fig2}
\end{figure}

Magnetic order of the Ir spins was detected by zero-field (ZF)
muon-spin rotation ($\mu^{+}$SR) on a powder sample of
Na$_2$IrO$_3$. Example raw spectra are shown in Fig.~\ref{fig2}a).
At temperatures below $T_{N}=15.3$~K, we observe clear
oscillations in the time-dependence of the muon polarization,
characteristic of quasi-static local magnetic fields at the muon
stopping site. Fits to the time-dependent muon data reveal that
two frequencies are present, indicating the presence of two
distinct muon stopping sites with different local fields. The full
spectra was fitted to the form $A(t)  =
A_{1}e^{-\lambda_{1}t}\cos(2\pi\nu_{1}t+\phi_{1})+ A_{2}
e^{-\lambda_{2}t}\cos(2\pi\nu_{2}t+\phi_{2})+ A_{3}e^{-\Lambda
t}+A_{\mathrm{bg}}$, where the last two terms account for muons
polarized parallel to the local magnetic fields, and muons
stopping in the sample holder (or cryostat tail), respectively.
Using our best-fit parameters we estimate that the muons occupy
the two sites with a probability ratio of about 9:1. Both local
fields set in at a common temperature, but have a distinctly
different temperature dependence [see Fig.~\ref{fig2}b)]. The
relative weight of the second frequency component suggests that it
may come from muon sites implanted near stacking fault planes, as
such sites also occur in a similar proportion. Our value for
$T_{\rm N}$ is consistent with both susceptibility measurements on
the same batch, which indicated a clear anomaly (sharp downturn)
near $T_{N}$ as reported previously \cite{gegenwart,Yogesh}, and
the magnetic Bragg peaks observed in resonant xray scattering
\cite{hill}.

\begin{figure}[t!]
\begin{center}$
\begin{array}{c}
\includegraphics[width=8.5cm]{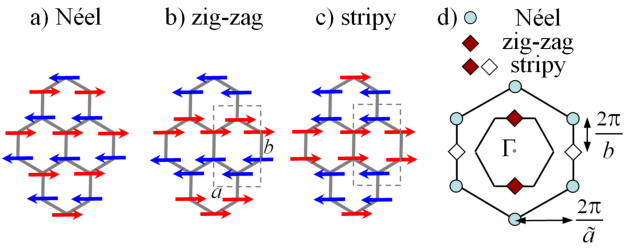} \\
\includegraphics[width=8.5cm]{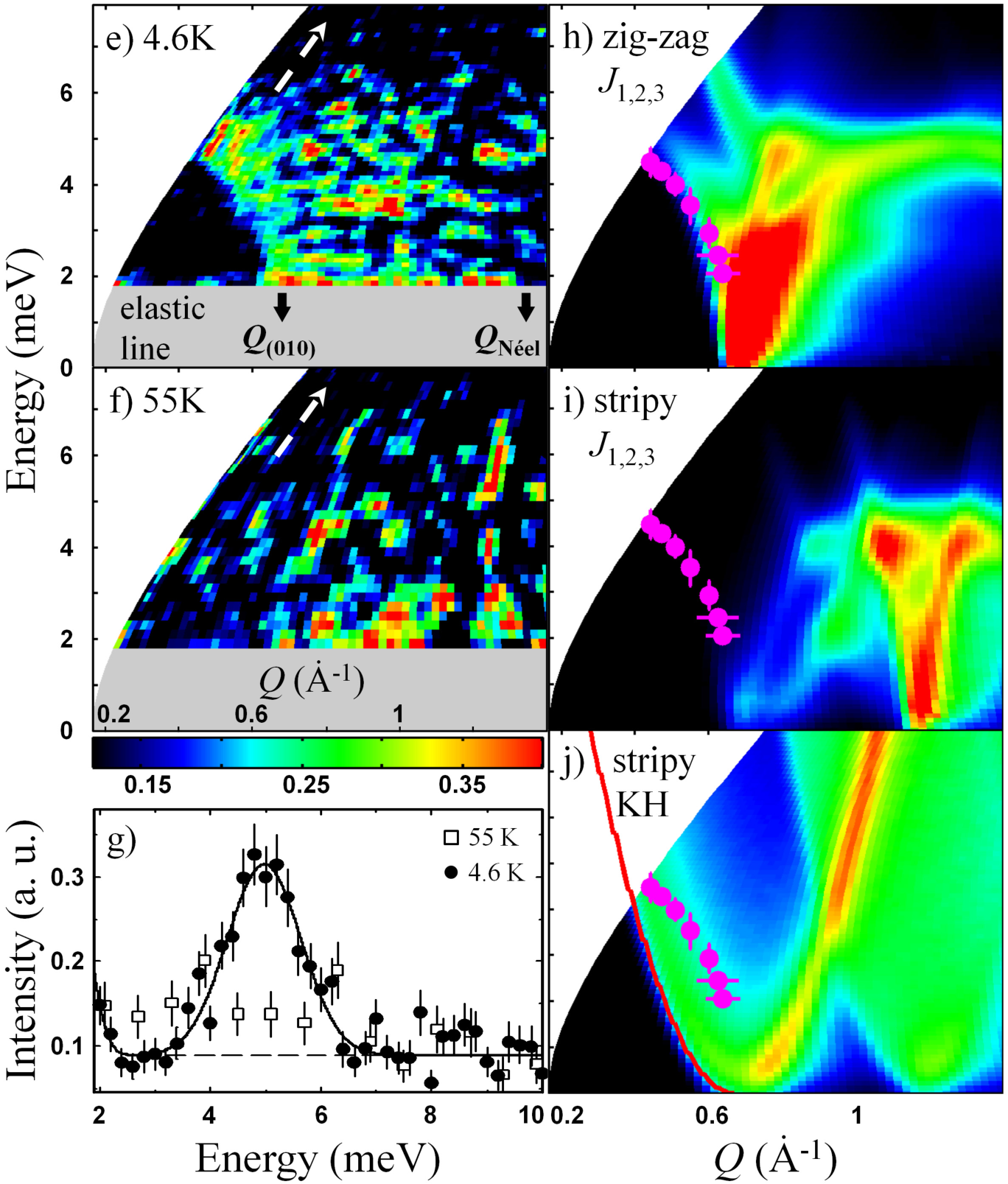}
\end{array}$
\end{center}
\vspace{-0.3cm} \caption{(Color online) Diagram of a) N\'{e}el, b)
zig-zag and c) stripy order. d) Reciprocal space diagram showing
locations of magnetic Bragg peaks for various magnetic phases
(inner hexagon shows first Brillouin zone of the honeycomb
lattice, $\tilde{a}=a\sin\beta$.) e) Powder inelastic neutron scattering data. The notable
well-defined feature is the sharp lower boundary of the scattering
at low $Q$ (filled (magenta) symbols in h-j)), which we associate
with a sinusoidal spin wave dispersion; this becomes damped out in
the paramagnetic phase in f). Slanted thick dashed arrow shows the
scan direction in g). Gray shading marks the inaccessible region
close to the elastic line dominated by incoherent elastic
scattering. g) Energy scan (solid points 4.6~K, open symbols 55~K)
through the maximum spin-wave energy seen in e) fitted to a
Gaussian peak (solid line), dashed line is estimated background.
h-j) Calculated spherically-averaged spin-wave intensity
\cite{supplementary} for the $J_{1,2,3}$ model with h) zig-zag or
i) stripy order, and j) the KH model with stripy order for
parameters given in the text. Solid red line in j) highlights the
low-energy boundary, which coincides with the dispersion from
$\Gamma$ to the first softening point.} \label{ins}
\end{figure}

The magnetic excitations were probed by powder inelastic neutron
scattering using the direct-geometry time-of-flight spectrometer
MARI at ISIS with an optimised setup to minimise absorption
\cite{supplementary}. Fig.~\ref{ins}e) shows the raw neutron
scattering intensity as a function of wavevector ($Q=|\bm{Q}|$)
and energy transfer deep in the ordered phase. An inelastic signal
with a sinusoidal-like dispersive boundary below a maximum near 5
meV is clearly observed at low $Q$. A gap, if present is smaller
than 2 meV. The magnetic character of the scattering is confirmed
by the broad, damped-out signal observed in the paramagnetic phase
at 55 K [see Fig.~\ref{ins}f) and g) (contrast filled and open
symbols)]. Interestingly, the dispersion boundary extrapolates at
the lowest energies to a wavevector $Q$ much smaller than that
expected for conventional N\'{e}el order, $Q_{(020)}= 1.34$ \AA
$^{-1}$, so this magnetic order can be ruled out; in fact $Q$ is
close to the expected location of the first magnetic Bragg peak
for both zig-zag or stripy order, $Q_{(010)}=0.67$~\AA $^{-1}$.
Figs.~\ref{ins}h) and i) show the calculated scattering from spin
waves of a 2D Heisenberg model with up to 3rd neighbour exchanges,
$J_{1,2,3}$, with zig-zag ($J_1=4.17$\ meV, $J_2/J_1=0.78$,
$J_3/J_1=0.9$) and stripy order ($J_1=10.89$\ meV, $J_2/J_1=0.26$,
$J_3/J_1=-0.2$), respectively (we neglect the interlayer couplings
believed to be small). The constraints to reproduce the dispersion
maximum and the measured Curie-Weiss (CW) temperature
($\Theta=-S(S+1)(J_{1}+2J_{2}+J_{3})/k_{\mathrm{B}} \sim-125$~K
\cite{Yogesh}) are not sufficient to determine all 3 exchanges, so
the values chosen are {\em representative} of the level of
agreement that can be obtained \cite{supplementary}. The
calculation for the zig-zag phase [Fig.~\ref{ins}h)] can reproduce
well the observed dispersion at low-$Q$ (filled symbols), whereas
the stripy phase [Fig.~\ref{ins}i)] cannot account for the strong
low-$Q$ dispersive signal and predicts stronger scattering at
larger-$Q$'s not seen. Calculations for the KH Hamiltonian
(\ref{eq_HK}) are shown in Fig.~\ref{ins}j) for $\alpha=0.4$
(lower limit for the stripy phase) and $J_1=25.85$~meV to
reproduce the CW temperature \cite{trebst_chi}
$\Theta=-S(S+1)(J_1-J_{\mathrm{K}}/3)/k_{\mathrm{B}}$. The lower
boundary of the scattering at low $Q$ (solid line) is predicted to
have a quadratic shape near the first softening point, a robust
feature for any $\alpha$ throughout the stripy phase. This is in
contrast to the data where the dispersion boundary (marked by
filled symbols) has a distinctly different, sinusoidal-like shape
with a curvature the opposite way. In addition, a different
distribution of scattering weight to higher energies is predicted,
but not seen in the data. We conclude that the KH model in the
stripy phase has a qualitatively different spin-wave spectrum
compared to the data. A minimal model that can reproduce the
observed low-$Q$ dispersion and which predicts distribution of
magnetic scattering in broad overall agreement with the data up to
some intensity modulations is shown in Fig.~\ref{ins}h) and
requires substantial couplings up to 3rd neighbors, which
stabilize zig-zag magnetic order. Recent theory \cite{Kimchi}
proposed that {\em in addition} to couplings up to 3rd neighbors,
a Kitaev term may also exist. We have compared the data with such
a model as well \cite{supplementary} and estimate that a Kitaev
term, if present, is smaller than an upper bound corresponding to
$\alpha\lesssim 0.40(5)$.

We note that sizeable $J_{3}$'s are not uncommon in triangular
plane metal oxides. The reason is that even though $J_{1}$
involves two hoppings and $J_{3}$ four, the two additional
hoppings are strong $pd\sigma$ ones, and the hopping proceeds
through intermediate unoccupied $e_{g}$ states \cite{NiGa}. In
case of Na$_{2}$IrO$_{3}$ the hopping proceeds through somewhat
higher Na $s$ orbitals, but these are very diffuse, and the
corresponding $t_{sp\sigma}$ parameter is sizeable. Near
cancellation of the AFM and FM superexchange interaction for the
nearest-neighbor path further reduces $J_{1}$ compared to $J_{3}$.

To summarize, by combining single-crystal diffraction and LDA
calculations we proposed a revised crystal structure for the
spin-orbit coupled honeycomb antiferromagnet Na$_2$IrO$_3$ that
highlights important departures from the ideal case where the
Kitaev exchange dominates. We observed dispersive spin-wave
excitations in inelastic neutron scattering and showed that
substantial further-neighbor exchange couplings are required to
explain the observed dispersion and we proposed a model for the
magnetic ground state that could support such a dispersion
relation.

We thank G.~Jackeli for providing notes on spin-wave dispersions
for the KH model in the rotated frame, A.~Amato for technical
support, N.~Shannon, J.T. Chalker and L.~Balents for discussions,
and EPSRC for funding. Work at Rutgers was supported by DOE
(DE-FG02-07ER46382).

\vspace{-0.2cm}



\vspace{1.7cm}

\textbf{ \em{Supplemental Material for}} \textbf {Spin waves and
revised crystal structure of honeycomb iridate Na$_{2}$IrO$_{3}$}

\vspace{3 mm}

Here we provide additional information on 1) structural
optimization calculations to confirm the unit cell stability and
estimate the energy of stacking faults, 2-3) the xray diffraction
measurements and analysis of the diffuse scattering, 4) $\mu$SR
and 5) neutron scattering experiments, and 6-9) derive the
spin-wave dispersion relations and dynamical structure factor in
neutron scattering for the Heisenberg $J_{1,2,3}$,
Kitaev-Heisenberg and Kitaev-Heisenberg-$J_2$-$J_3$ models for
various magnetic orders.

\maketitle

\vspace{5 mm}

\renewcommand{\thefigure}{S1}
\begin{figure}[t!]
\begin{center}
\includegraphics[width=\linewidth,angle=0]{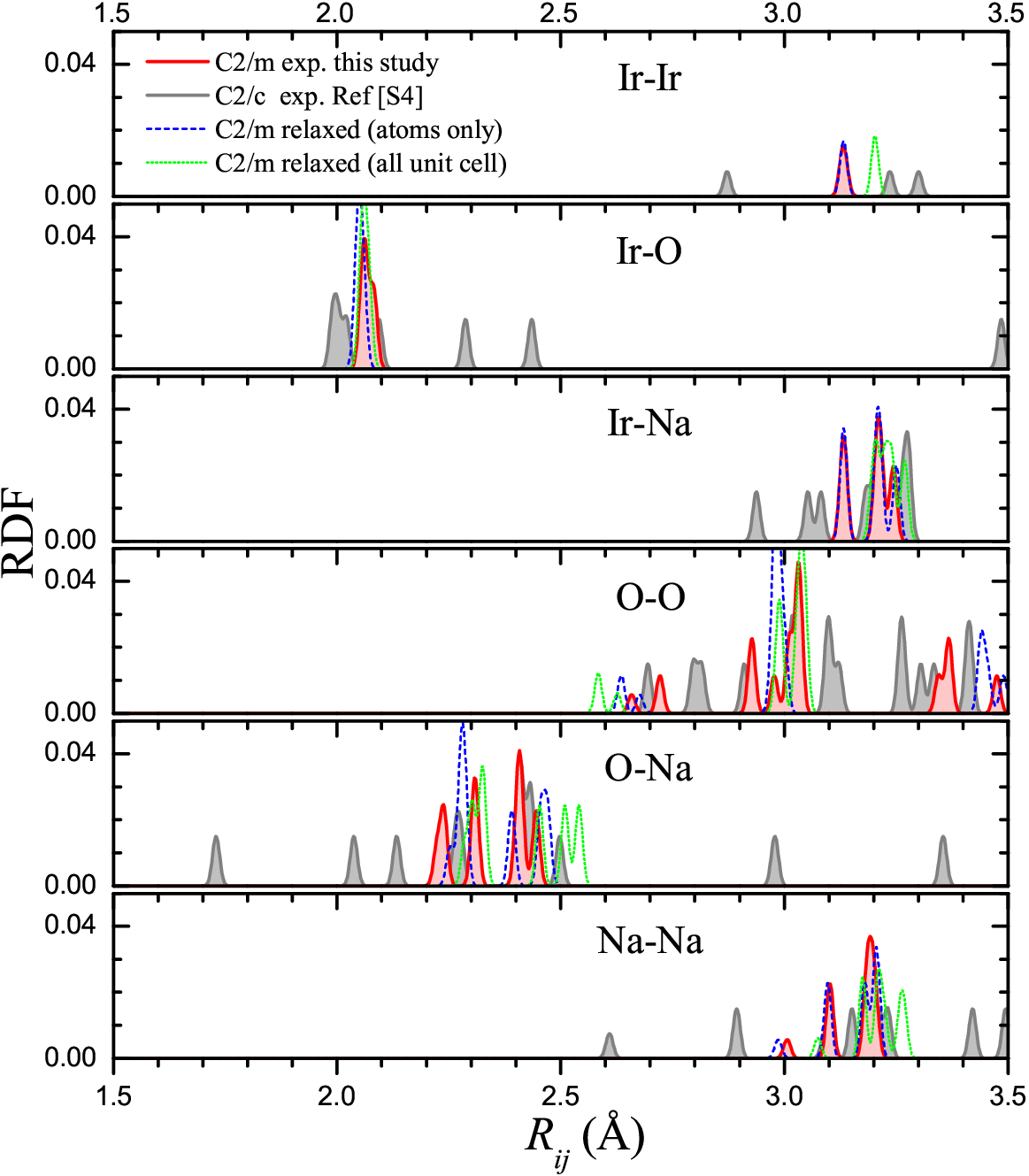}
\caption{(Color online) Radial distribution functions (RDFs)
showing the difference in the local atomic environments for the
previously reported $C2/c$ structure \cite{S_gegenwart} and the
$C2/m$ structure proposed in this study. In the $C2/m$ case, the
RDFs are plotted for the unit cell extracted from the experiment
(red solid line), the unit cell with atomic positions relaxed in
the DFT (blue dashed line), and the unit cell fully relaxed in the
DFT (green dotted line). Note that a small Gaussian smearing
($\sigma = 0.008$\AA ) was used in the calculation of the
RDFs.}\label{fig_rdf}
\end{center}
\end{figure}

{\bf S1. Structural optimization calculations using {\small VASP}}\\

We used the Perdew-Burke-Ernzerhof (PBE) exchange-correlation
functional \cite{S_PBE} within the generalized gradient
approximation (GGA) and the projector augmented waves method
\cite{S_PAW}. The $2p$ semi-core electrons in Na were treated as
valence. Numerical convergence was achieved with a 500 eV energy
cutoff and dense Monkhorst-Pack $k$-meshes \cite{S_MONKHORST_PACK}
of 7$\times$7$\times$3 for the previously reported
\cite{S_gegenwart} $C2/c$ primitive unit cell and
6$\times$4$\times$6 for the proposed $C2/m$ conventional unit cell
in Table I. We performed three types of calculations for the two
structures: a static run with the experimental parameters,
optimization of the atomic positions only, and full optimization
of the atomic positions and lattice parameters. The residual
forces and stresses were typically below 0.002 eV/\AA \ and 0.5
kbar, respectively. We found the magnetic and the spin-orbit
interactions to have a rather small effect on the Na$_2$IrO$_3$
structure and the comparisons below are made for the non-magnetic
case without the spin-orbit coupling.\\

To illustrate the differences in the local environments in Fig.\
\ref{fig_rdf} we plotted normalized radial distribution functions
(RDFs) for all types of interatomic distances in the experimental
and optimized structures. $C2/c$ exhibits a considerable
dispersion of the Ir-Ir and Ir-O nearest neighbor distances
critical for the magnetic ordering in the compound. The O-Na and
Na-Na separations are unphysically small and we observed large
forces, over 6 eV/\AA \ on Na and over 3 eV/\AA \ on O, at the
beginning of the optimization run. The RDFs in $C2/m$ with the
experimental parameters demonstrate much more symmetric local
environments and a negligible variation of Ir-Ir lengths within
the honeycomb lattice (below 0.3\%). The calculated forces on
atoms did not exceed 0.5 eV/\AA \ indicating a good agreement
between the experiment and theory. Optimization of the atomic
positions with fixed $C2/m$ experimental unit cell had little
effect on the Ir-Ir distances because they are defined primarily
by the in-plane lattice constants $a$ and $b$. When fully
optimized, $C2/c$ and $C2/m$ converged to the same structure with
the $C2/m$ space group and virtually indistinguishable RDFs. The
enthalpy gains were 0.434 and 0.018 eV/atom, respectively (for
comparison, the optimization of atomic positions in $C2/m$ led to
a 0.007 eV/atom gain). Note that the full optimization of $C2/m$
leads to $\sim$ 2\% elongation of the Ir-Ir distances which is a
typical bond overestimation observed for the GGA. For this reason
we believe that use of the experimental lattice constants is more
appropriate for the modelling of the magnetic interactions.

To estimate the stacking fault energy we simulated
1$\times$1$\times$$n$ ($n=2,\dots,6$) supercells of the $C2/m$
primitive 12-atom unit cell with one Ir-Na layer and the two
adjacent O layers shifted by $b/3$ along [010]. The resulting
lower-symmetry structures ($C2$ space group) had two stacking
faults per unit cell and the same $a\times b/2=25.49 $\AA$^2$
$x-y$ base. We optimized only the atomic positions keeping the
experimental unit cell parameters fixed. The $n=2$ structure
gained additional symmetry operations ($C2/c$ space group) upon
relaxation. The comparison of the faulted structures against the
respective $C2/m$ supercells with the same unit cell dimensions
and the same $k$-point meshes allowed us to reduce computational
errors. However, the energy differences, $E_{n}-E_{C2/m}$, in our
non-magnetic calculations without the spin-orbit coupling (SOC)
proved to be exceptionally small in magnitude: 0.7, -1.7, -2.0,
-2.6, -1.8 meV/($n\times 12 $ atoms) for $n=2,\dots,6$,
respectively. For the smallest $n=2$ structure we were able to
calculate the energy difference with the FM ordering and the SOC
as well and found $E_{n=2}-E_{C2/m}$ to remain small at 2.9
meV/(24 atoms). Based on these tests, we expect the stacking fault
energy in $C2/m$ to be below $\sim$ 0.1 meV/\AA$^2$, one to two
orders of magnitude smaller than typical stacking fault energies
for elemental metals. For comparison, an ABCBA stacking fault
generated by reflecting $C2/m$ structure (which has the ABCABC
sequence along $c$) about a Na layer was calculated to have a much
higher, measurable energy value of about 8 meV/\AA$^2$.\\

{\bf S2. Xray diffraction and structural analysis}\\

\renewcommand{\thefigure}{S2}
\begin{figure}[t!]
\begin{center}
\includegraphics[width=8.5cm]{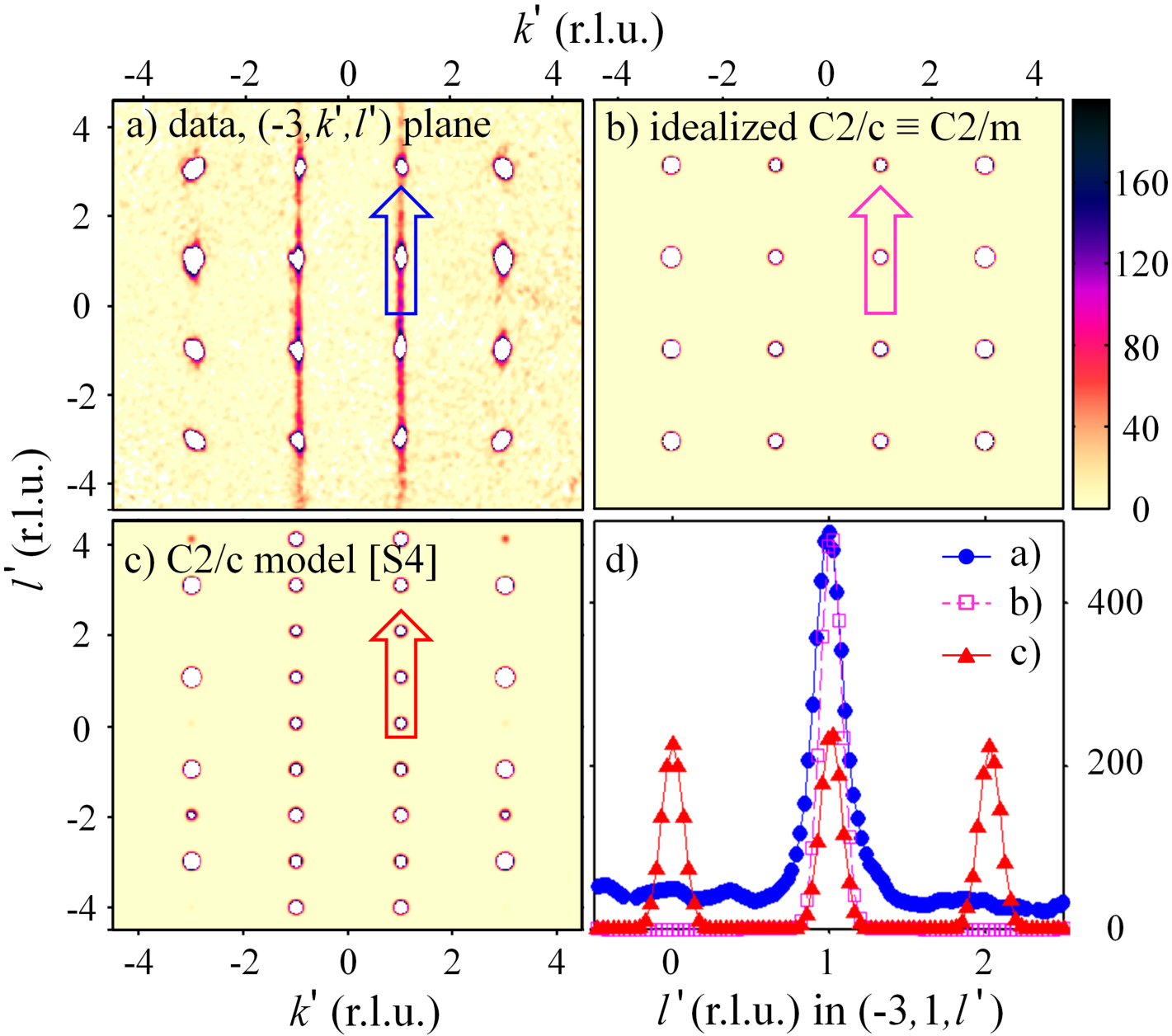}
\caption{(Color online) Xray diffraction intensity in the
($-3,k',l'$) plane: a) data, b) calculation for the ``idealized''
C$2/c$ structure with atoms at special positions, equivalent to
the C$2/m$ model in Table I, c) calculation for the distorted
C$2/c$ model in \cite{S_gegenwart} (assuming no Ir/Na site mixing
in the honeycomb Ir$_{2/3}$Na$_{1/3}$ layers). Notice the series
of sharp peaks predicted in c) at (-3,1,even) positions, which
however are not present in the data in a). d) Scan along the
(-3,1,$l'$) line (arrowed direction in a-c)) comparing data (solid
circles) and calculation for the two models (triangles-C$2/c$ and
squares-C$2/m$). The calculated diffraction intensities have been
multiplied by an overall scale factor and have been convolved with
a finite width Gaussian in momentum space to mimic the effects of
the instrumental resolution.}\label{figS2_xray}
\end{center}
\end{figure}

X-ray diffraction was performed using a Mo-source Oxford
Diffraction Supernova diffractometer on a single crystal of
Na$_{2}$IrO$_{3}$ of approximate size
220$\times$150$\times$10$\mu$m$^{3}$ grown via flux
\cite{S_gegenwart}. 96$\%$ out of over 1000 detected peaks were
indexed by a single monoclinic domain. Structural refinement was
performed using both a unit cell with space group C$2/m$, with
parameters listed in Table I, as well as a unit cell with twice
the volume and space group C$2/c$, using the SIR-92 and SHELX
packages \cite{S_single_ref}. The two unit cell parameters are
related by ${\bm a^{\prime}}=-\bm{a}$, ${\bm b^{\prime}
}=-\bm{b}$, ${\bm c^{\prime}}=\bm{a}+2\bm{c}$,
$c^{\prime}=\sqrt{a^{2} +4c^{2}+4ac\cos\beta}$,
$\sin\beta^{\prime}=\frac{2c}{c^{\prime}}\sin\beta$, and in terms
of the reciprocal lattice components $h^{\prime}=-h$, $k^{\prime
}=-k$, $l^{\prime}=h+2l$, where primed values refer to the C$2/c$
model. Starting from the larger unit cell (C$2/c$) and slightly
displacing the atoms to some ``ideal'' positions one recovers the
higher-symmetry structure described by the smaller, C$2/m$, cell.
The distinction between those two models is entirely due to such
small atomic displacements, the presence of which is manifested in
finite intensity diffraction peaks at ($h',k',l'$) positions with
$h'$ odd, $k'$ odd and $l'$ even, which disappear when atoms are
displaced to the ``ideal'' positions, when the structure recovers
the C$2/m$ symmetry. This is illustrated by the calculated
diffraction pattern in the ($-3,k',l'$) plane where the ``extra''
peaks expected in the larger cell model C$2/c$ shown in Fig.\
\ref{figS2_xray}c) are not seen in the data plotted in Fig.\
\ref{figS2_xray}a), which is however fully consistent with the
pattern expected for the higher-symmetry C$2/m$ model shown in
Fig.\ \ref{figS2_xray}b). This is also apparent in Fig.\
\ref{figS2_xray}d) showing a scan along the ($-3,1,l'$) line with
extra peaks (triangles) predicted for $l'=0,2$, not seen in the
data (filled symbols). For completeness we note that we applied a
shift of the fractional atomic coordinates in the C$2/c$ unit cell
(in the notation adopted in \cite{S_gegenwart}) by (-1/4,-3/4,0)
before converting them into fractional atomic coordinates of the
C$2/m$ cell (in the notation used in Table I), due to the
different positions of the origin in the two space
groups.\\

{\bf S3. Microscopic model of stacking faults}\\

The calculated diffraction pattern in Fig.\ 1e) was obtained
numerically by direct structure-factor calculations using the
DISCUS package \cite{S_DISCUS}. We considered a ``crystal'' of
$200 a \times 200 b\times 4000 c$ unit cells of Na$_2$IrO$_3$
(C$2/m$). To include the effect of stacking faults we assumed that
each Ir layer has a choice with probability $1-p$ to keep
in-stacking-sequence with the layer below and $p/2$ to be shifted
to either of the other two sublattice positions (translated
in-plane by $(0,1/3,0)$ or $(1/2,1/6,0)$), with $p=0$ for perfect
stacking and $p=1/3$ for a completely uncorrelated layer stacking
sequence, a model first introduced to describe the stacking faults
in the related material Li$_2$MnO$_3$ \cite{S_Li2MnO3}.\\

{\bf S4. Muon spin relaxation experiments}\\

Zero field (ZF) $\mu^{+}$SR measurements were made at the Swiss
Muon Source (S$\mu^{+}$S), Paul Scherrer Institut, CH using the
GPS spectrometer. For the measurement a 250~mg powder sample of
Na$_{2}$IrO$_{3}$, which was used for inelastic neutron scattering
measurement, was packed inside a silver foil packet (foil
thickness 25$~\mu$m) and mounted on a silver sample holder.

Fits of the data to an equation in main text reveal the evolution
of $\nu_{i}$ and $\lambda_{i}$ with temperature, as shown in Figs.
2(b-c). Unusually, the frequencies do not vary in fixed
proportion, although they do tend to zero at the same temperature.
The low-amplitude, higher frequency component $\nu_{2}$ drops off
far more dramatically than the large amplitude, lower frequency
$\nu_{1}$. In order to quantify this behavior, the frequencies
were fitted to the phenomenological function $\nu_{i}(T) =
\nu_{i}(0) \left[1 - \left( T/T_{\mathrm{N}}\right)^{\alpha_{i}}
\right]^{\beta_{i}}$. A common value of $T_{\mathrm{N}}=15.3(1)$\
K was identified from fitting to this function. We find that
$\alpha \approx 2$ for both cases. The parameter $\beta$ can be
interpreted as an order parameter exponent. The other fit
parameters are $\nu_{1}(0)=5.54(1)$~MHz, $\beta_{1}=0.36(1)$,
$\nu_{2}(0)=6.20(3)$~MHz and $\beta_{2}=0.11(1)$. We note that
$\lambda_{1}$ is an order of magnitude larger than $\lambda_{2}$,
implying either that the distribution of fields is broader in the
majority site or, assuming the fast fluctuation limit, that the
fluctuation rate is smaller. The lower frequency oscillation,
accounting for $\approx 90$\% of the muon sites in the material,
has a $\beta$ value suggestive of the behavior of a
three-dimensional (3D) system (for 3D Heisenberg $\beta=0.367$ and
3D Ising $\beta=0.326$), while the minority muon site has an
exponent value more similar to that expected for a 2D Ising system
(for which $\beta=0.125$). These seem to suggest that the magnetic
fluctuations have a rather
different character at the two muon sites.\\

{\bf S5. Inelastic neutron scattering experiments}\\

Inelastic neutron scattering measurements were made using the
direct-geometry time-of-flight spectrometer MARI at ISIS using an
incident neutron energy of 18 meV, which covered the full
bandwidth of magnetic excitations with a zone boundary energy near
5 meV. The instrumental energy resolution was 0.67(1)~meV (FWHM)
on the elastic line. The sample was $\sim10$~g of Na$_2$IrO$_3$
powder spread out in a very thin layer ($\lesssim 1$~mm to
minimise neutron absorption) inside of an annular can of outer
diameter of 40 mm and height 50 mm. Counting times for the data in
Figs.\ 3e-f) were 28 and 7 hours, respectively, at an average
proton current of $150\mu$Amps.\\

{\bf S6. Spin-wave dispersions for the Heisenberg $J_{1,2,3}$
model in the zig-zag and stripy phases} \\

Here we outline the derivation of the linear spin wave dispersion
relations and dynamical structure factors relevant for neutron
scattering for various spin Hamiltonians on the honeycomb lattice.
For the Heisenberg model with up to 3rd neighbour exchanges we
extend previous results on the dispersion relations
\cite{S_Rastelli} to include also the dynamical structure factors.
For the Kitaev-Heisenberg model the spin-wave spectrum (including
$1/$S quantum corrections) has been studied before in a special
``rotated'' reference frame \cite{S_jackeli}, here we explicitly
derive here the dispersion relations and dynamical structure
factors in the experimentally-relevant, un-rotated reference
frame. For the Kitaev-Heisenberg-$J_2$-$J_3$ models both the
dispersion relations and dynamical structure factors have not been
studied before.

\renewcommand{\thefigure}{S3}
\begin{figure*}[t!]
\begin{center}
\includegraphics[width=17cm]{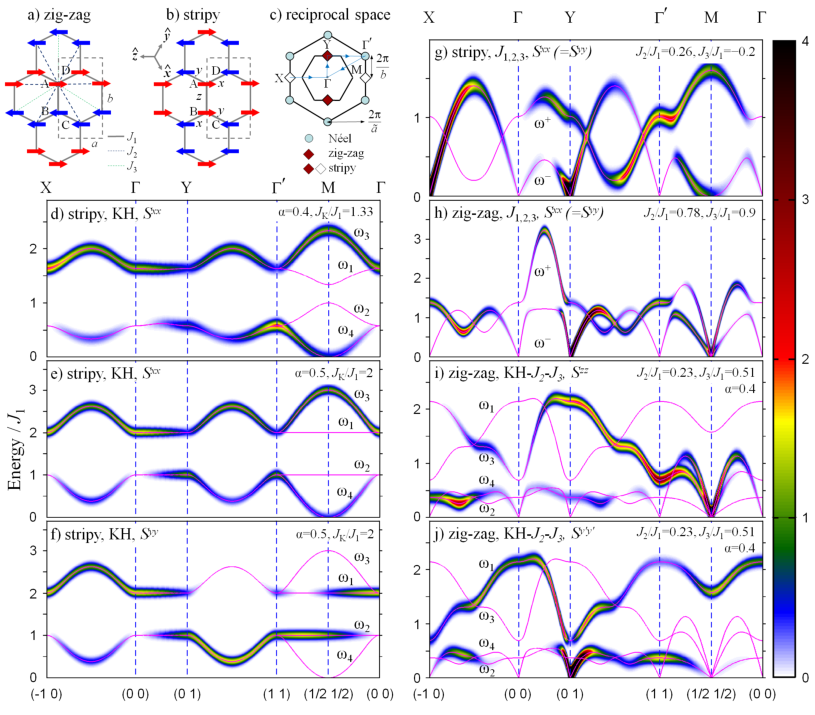}
\caption{(Color online) a) Zig-zag and b) stripy order. Dashed
rectangular box of size $a \times b$ shows the magnetic unit cell
containing four sublattices A-D. In a) solid, dashed and
dot-dashed lines show paths for $J_1$, $J_2$ and $J_3$. b) Bond
labels $x,y,z$ refer to the components of the spins at the two
bond ends coupled by the Kitaev term. Inset show projection of the
(cubic) $\hat{\bm x}$, $\hat{\bm y}$ and $\hat{\bm z}$ axes onto
the honeycomb plane. c) 2D reciprocal space showing magnetic Bragg
peak positions for various magnetic orders, $\tilde{a}=a\sin\beta$.
d-j) Spin-wave dispersions along symmetry directions in reciprocal space (arrowed
path in c)) for the KH, $J_{1,2,3}$ and KH-$J_2$-$J_3$
Hamiltonians for exchange values and magnetic orders listed in the
legends. Wavevectors $\bm{Q}$ are expressed in reciprocal lattice
units of the rectangular magnetic unit cell. Colour is the
dynamical structure factor (convolved with a Gaussian in energy
for visualization, full width at half maximum~$=0.15J_1$),
isotropic for the Heisenberg model in g-h)
($S^{xx}(\bm{Q},\omega)=S^{yy}(\bm{Q},\omega)$) and different for
the two polarizations $x, y$ for the KH model in d-f). i-j)
Dynamical structure factor for the KH-$J_2$-$J_3$ model with the
zig-zag structure in a), where the ordered moments are along a
general direction $x'$ in the $xy$ plane and $y'$ is a direction
in this plane normal to $x'$.}\label{fig_disp}
\end{center}
\end{figure*}

We start with the isotropic Heisenberg model on the honeycomb
lattice with exchanges with up to 3rd nearest-neighbor, so called
$J_{1,2,3}$ model with Hamiltonian
\begin{equation} {\cal H}=\sum_{\rm{1NN}} J_1 {\bm S}_i \cdot
\bm{S}_j+\sum_{\rm{2NN}}J_2 {\bm S}_i \cdot \bm{S}_k
+\sum_{\rm{3NN}}J_3 {\bm S}_i \cdot \bm{S}_l  \tag{S1}
\label{eq_J123}
\end{equation}
where 1-, 2-, and 3NN indicate summing over all 1st, 2nd and 3rd
nearest-neighbor pairs with couplings $J_1$, $J_2$ and $J_3$
[paths indicated in Fig.\ \ref{fig_disp}a)], where positive values
correspond to antiferromagnetic exchanges. Depending on the
relative ratio of the couplings there are six distinct types of
mean-field ground states \cite{S_Rastelli,S_fouet}, which include
the two candidate magnetic orders for Na$_2$IrO$_3$, the zig-zag
and stripy AFM orders shown in Figs.\ \ref{fig_disp}a-b) (labelled
II and IV, respectively, in \cite{ S_Rastelli,S_fouet}). Both of
those magnetic structures have four magnetic sublattices (labelled
A-D) and can be described by a rectangular magnetic unit cell
(dashed box in Figs.\ \ref{fig_disp}a-b)), which coincides with
the in-plane chemical unit cell $a \times b$ of Na$_2$IrO$_3$.
Within a single layer the Ir honeycomb lattice in very close to
ideal ($b/a\simeq\sqrt{3}$) in spite of the 3D monoclinic crystal
structure, so we treat here the ideal 2D honeycomb lattice with
3-fold symmetry. In this case the magnetic order can have three
spacial domains, one such domain is shown for both structures in
Figs.\ \ref{fig_disp}a-b), the other two magnetic domains are
obtained by $\pm 60^{\circ}$ rotation around the direction normal
to the plane.

Using a standard Holstein-Primakoff transformation in the
large-$S$ limit the Hamiltonian becomes (to leading order) a
quadratic form of magnon operators
\begin{equation}
\mathcal{H}=\sum_{\bm  q}\mathsf{X}^{\dagger}\mathsf{H}\mathsf{X}
+N(1+1/S)E_{MF} \tag{S2} \label{quadratic_ham}
\end{equation}
where higher than quadratic terms are neglected. Here $E_{MF}$ is
the mean-field ground state energy (per spin) and $N$ is the total
number of spin sites. The sum extends over all wavevectors ${\bm
q}$ in the first magnetic Brillouin zone.

For the zig-zag order in Fig.\ \ref{fig_disp}a) we define the
operator basis as $\mathsf{X^\dag}=\left[a^\dagger_{\bm q} ~,~
d^\dagger_{\bm q} ~,~ c_{-\bm q} ~,~ b_{-\bm q}\right]$ where
$a-d$ label operators on sublattice A-D, i.e. $a^\dagger_{\bm q}$
($a_{\bm q}$) creates (annihilates) a plane-wave magnon mode on
sublattice A and so on. The Hamiltonian matrix in eq.\
(\ref{quadratic_ham}) is
\begin{equation}
\mathsf{H}=\left[\begin{array}{llll}
A&   B&  C & D^* \\
B^*&  A&  D& C \\
C& D^* & A& B \\
D& C & B^* & A \\
\end{array}\right] \tag{S3} \label{Ham_matrix}
\end{equation}
where
\begin{displaymath}
\begin{array}{l l}
    A&=S\ \{-J_1+2J_2+3J_3+2J_2\cos(2\pi h) \} \\
    B&=2SJ_1\eta\cos(\pi h) \\
    C&=2SJ_2\{\cos\left[\pi(h+k)\right]+\cos\left[\pi(h-k)\right]\} \\
    D&=S\ \{J_1{\eta}^{2}+J_3\left[{{\eta}}^{-4} +2{{\eta}}^{2}\cos(2\pi h)\right]\}\\
    \eta&=e^{k \pi i/3}.
\end{array}
\end{displaymath}
Here $(h,k)$ are components of the wavevector ${\bm q}$ in units
of the reciprocal lattice of the $a \times b$ rectangular unit
cell shown in Fig.\ \ref{fig_disp}a). By periodicity the above
expressions are valid for any momentum, not necessarily restricted
to the 1st magnetic Brillouin zone. Diagonalisation of the
Hamiltonian by standard techniques \cite{S_white, S_WheelerAgNiO2}
to obtain the normal magnon modes gives two doubly-degenerate
dispersions
\begin{equation}
\begin{array}{ll}
(\omega^{\pm}_{\bm q})^2&=A^2+BB^*-C^2-DD^* \\
&\pm\sqrt{4|AB-CD^*|^2-|B^*D^*-BD|^2}.
\end{array} \tag{S4}
\label{HB_disp}
\end{equation}

We have explicitly verified for the same model (\ref{eq_J123})
that eq. (\ref{HB_disp}) agrees with
earlier results of \cite{S_Rastelli} [eq. (5.21)]. 
The spin-wave intensity in neutron scattering is proportional to
the dynamical structure factor (expressed as $S^{xx}({\bm
Q},\omega)$ for spin fluctuations along the $x$-direction and
similarly for $y$-direction) and an analytical expression for this
in the case of a Hamiltonian of the form in eq.\
(\ref{Ham_matrix}) are given explicitly \cite{S_WheelerAgNiO2}
[eq. (A3)] and for brevity are not included here.

The spin-wave dispersions in (\ref{HB_disp}) (and their intensity
dependence) for the zig-zag phase are plotted for representative
exchange values in Fig.\ \ref{fig_disp}h). As expected, the
acoustic magnon, $\omega^-$, is gapless with a linear dispersion
at the magnetic Bragg peak at the Y point, is also linear and
gapless at the X point, but has zero intensity because the
structure factor for magnetic Bragg peaks also cancels at this
position. Furthermore, the dispersions soften and appear gapless
at the M point and others part of the quartet ($\pm 1/2,\pm 1/2$),
which are Bragg peaks for the other two magnetic domains rotated
by $\pm60^{\circ}$. This softening is a general feature of linear
spin-wave dispersions for a multi-domain magnetic ground state
\cite{S_WheelerAgNiO2}, however the fact that the gap closes at
those points is not protected by any symmetry, but is an artefact
of the linear spin-wave approximation; by analogy with related
spin-wave models for other multi-domain structures
\cite{S_chubukov} we expect the dispersions to become gapped at
the softening points when quantum fluctuations to 1st order in
1/$S$ are included.

\renewcommand{\thefigure}{S4}
\begin{figure}[tb]
\begin{center}
\includegraphics[width=8.5cm,angle=0]{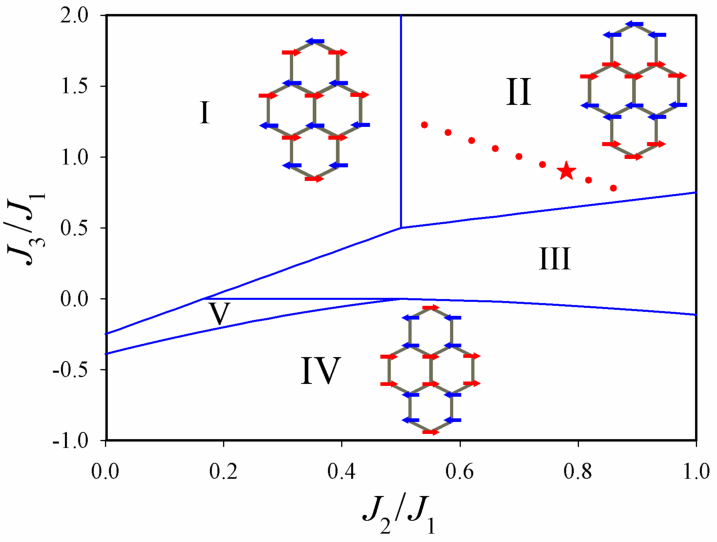}
\caption{(Color online) Classical phase diagram of the Heisenberg
$J_{1,2,3}$ model on the honeycomb lattice, eq.\ (\ref{eq_J123}),
showing the regions of stability for zig-zag (II) and stripy order
(IV). Phase I is collinear $2$-sublattice N\'{e}el order, III and
V are incommensurate spiral phases, and solid lines are phase
boundaries \cite{S_fouet}. The dotted line inside region II
indicates possible solutions for a minimal model to describe the
spin dynamics in Na$_2$IrO$_3$ obtained by imposing the
constraints described in the text (the red star is a
representative solution for which the full spectrum is shown in
Fig.\ \ref{fig_disp}h)).} \label{fig_J2J3}
\end{center}
\end{figure}

A spherical average of the spin-wave spectrum (including various
prefactors listed in eq.\ (\ref{intensity}) below) is shown in
Fig.\ 3h). The dominant contribution to the low-$Q$ dispersive
edge of the strong signal near the first softening point ($Q$=0.67
\AA$^{-1}$) is due to acoustic magnons on the $\omega^{-}_{\bm k}$
branch emerging out of the Y point and dispersing in the
Y$\rightarrow\Gamma$ direction [see Fig.\ \ref{fig_disp}h)] and
also magnons on the $\omega^{+}_{\bm k}$ branch emerging out of
the M-point and dispersing in the M$\rightarrow\Gamma$ direction.
To reproduce the observed low-$Q$ dispersion in the powder data we
have imposed the constraint that the zone-boundary energy of the
lowest branch on the $\Gamma$-Y line reproduces the observed
maximum of the low-$Q$ dispersion, i.e.
$\omega^{-}\left(0,\frac{1}{2}\right)=5$\ meV. This constraint
together with the condition that the exchanges reproduce the
observed Curie-Weiss constant $\theta=-S(S+1)(J_1+2J_2+J_3)/k_{\rm
B}=-125$\ K cannot determine all three exchange values $J_1$,
$J_2$ and $J_3$, but allow for a one-dimensional family of
solutions located on a curve in the parameter space $(J_2/J_1,
J_3/J_1)$ (the dotted line in region II in Fig.\ \ref{fig_J2J3}).
All sets of exchange values part of this family are broadly
consistent with the data. The level of agreement that can be
obtained is illustrated in Fig.\ 3h) for one representative
solution (red star in Fig.\ \ref{fig_J2J3}), chosen as it comes
closest to reproducing also the intensity distribution at the
lowest $Q$.

We now turn to the alternative magnetic structure, the stripy
order shown in Fig.\ \ref{fig_disp}b). If the spin-wave operator
basis is defined as $ \mathsf{X^\dag}=\left[a^\dagger_{\bm q} ~,~
b^\dagger_{\bm q} ~,~ c_{-\bm q} ~,~ d_{-\bm q}\right]$, then the
Hamiltonian reduces to the same form as in eqs.\
(\ref{quadratic_ham},\ref{Ham_matrix}), with magnon dispersions
given by eq.\ (\ref{HB_disp}), but where the expressions for the
$A-D$ parameters are
\begin{displaymath}
\begin{array}{l l}
    A&=S\ \{J_1+2J_2-3J_3+2J_2\cos(2\pi h) \} \\
    B&=S\ \{J_1\eta^{-2}+J_3\left[\eta^4 +2\eta^{-2}\cos(2\pi h)\right]\}\\
    C&=2SJ_2\{\cos\left[\pi(h+k)\right]+\cos\left[\pi(h-k)\right]\} \\
    D&=2SJ_1{\eta}^{-1}\cos(\pi h) \\
    \eta&=e^{k \pi i/3}.
\end{array}
\end{displaymath}
The resulting spin-wave dispersions and intensities for
representative exchange values are plotted in Fig.\
\ref{fig_disp}g). In contrast to the zig-zag phase, for the stripy
phase the acoustic magnon, $\omega^-$, is gapless, with a linear
dispersion and finite intensity at both the X and Y points, as
both are magnetic Bragg peaks with non-zero structure factor (X
four times stronger intensity than Y). Again, due to the three
domain structure there is softening of the dispersion with an
artificial gapless point at M, which is expected to become gapped
when quantum fluctuations beyond the linear spin-wave
approximation are included, as discussed earlier. A spherical
averaging of the spin-wave spectrum in Fig.\ \ref{fig_disp}g) is
shown in Fig.\ 3i), here the strongest signal at low energies is
due to scattering from acoustic magnons near the X-point ($Q=1.16$
\AA$^{-1}$) with weaker scattering from magnons near Y ($Q=0.67$
\AA$^{-1}$) and intensity decreasing rapidly for magnons with smaller momentum.\\

{\bf S7. Spin-wave dispersions for the Kitaev-Heisenberg model
in the stripy phase}\\

For the nearest-neighbor Kitaev-Heisenberg (KH) model in eg.\ (1)
the stripy phase in Fig.\ \ref{fig_disp}b) is the stable ground
state for $0.4 \lesssim \alpha \lesssim 0.8$, where
$\alpha=J_{\mathrm{K}}/\left(J_{\mathrm{K}}+2J_1\right)$
\cite{S_jackeli}. This ground state is exact at $\alpha=0.5$, when
upon rotation of the coordinate system at certain sites the
Hamiltonian (1) converts to that of a Heisenberg ferromagnet in a
rotated basis \cite{S_jackeli}.

For each of the three bonds coming out of a honeycomb lattice site
the Kitaev term $J_{\rm K}$ couples different spin components $x,
y, z$ expressed in terms of an orthogonal (cubic) reference frame.
This is oriented with the cubic [111] axis normal to the honeycomb
plane and the projections of the $\hat{\bm x}$, $\hat{\bm y}$ and
$\hat{\bm z}$ axes in the plane making $120^{\circ}$ as shown in
Fig.\ \ref{fig_disp}b) inset. Each bond is labelled with the type
of the spin component for the moments at the two bond ends coupled
by the Kitaev term, i.e. the $z$-bond AB stands for exchange
$-J_{\rm K}S^z_{A}S^z_{B}$ and $x$-bond AD stands for $-J_{\rm
K}S^x_{A}S^x_{D}$ and so on.

Due to the anisotropic nature of the Kitaev exchange more coupling
terms between magnon operators on the 4 different magnetic A - D
sublattices are generated as compared to the Heisenberg
$J_{1,2,3}$ model. Thus, one needs to use the full 8-term operator
basis
$\mathsf{X}^\dagger=\left[a_{\bm q}^\dagger ~,~ b_{\bm q}^\dagger ~,~ c_{\bm q}^\dagger ~,~ d_{\bm q}^\dagger \\
~,~ a_{-\bm q} ~,~ b_{-\bm q} ~,~ c_{-\bm q} ~,~ d_{-\bm
q}\right]$, for which the Hamiltonian expressed in magnon
operators to leading order still has the quadratic form
(\ref{quadratic_ham}) with the matrix $\mathsf{H}$ given by
\begin{equation}
\label{HK_matrix}
\mathsf{H}=\frac{1}{2}\left[\begin{array}{llllllll}
 A&   B&  0&   C&  0&   0&  0&    D  \\  
 B^*& A&  C^*& 0&  0&   0&  D^*&  0  \\  
 0&   C&  A&   B&  0&   D&  0&    0  \\  
 C^*& 0&  B^*& A&  D^*& 0&  0&    0  \\  
 0&   0&  0&   D&  A&   B&  0&    C  \\  
 0&   0&  D^*& 0&  B^*& A&  C^*&  0  \\  
 0&   D&  0&   0&  0&   C&  A&    B  \\  
 D^*& 0&  0&   0&  C^*& 0&  B^*&  A  \\  
\end{array}\right] \tag{S5}
\end{equation}
where
\begin{displaymath}
\begin{array}{l l}
    A&=S\left(J_1+J_{\rm{K}}\right) \\
    B&=SJ_1\eta^{-2} \\
    C&=-SJ_{\rm{K}} i\sin\left(\pi h\right)\eta \\
    D&=S\left(2J_1-J_{\rm{K}}\right)\cos\left(\pi h\right)\eta \\
    \eta&=e^{k \pi i/3}.\\
\end{array}
\end{displaymath}
Diagonalization to get the normal magnon modes \cite{S_white}
gives four dispersion relations
\begin{equation}
\begin{array}{ll}
\omega_{1,2}^2({\bm q})=A^2-DD^* +|B-C|^2 \\
\pm\sqrt{4A^2|B-C|^2-|D^*(B-C)-D(B^*-C^*)|^2}\\
\omega_{3,4}^2({\bm q})=A^2-DD^*+|C+B|^2 \\
\pm\sqrt{4A^2|B+C|^2-|D^*(B+C)-D(B^*+C^*)|^2}.\\
\end{array} \tag{S6}
\label{eq_App_dispersion}
\end{equation}

The dispersion curves are plotted for $\alpha=0.4$ in Fig.\
\ref{fig_disp}d) and $\alpha=0.5$ in Figs.\ \ref{fig_disp}(e-f),
where the colour represents the dynamical structure factor,
plotted separately for the spin fluctuations along $x$ and
$y$-axes, the presence of Kitaev bond directional exchanges make
those the dynamical structure factor non-equivalent. The structure
factors were obtained from the eigenvectors of the Hamiltonian
matrix $\mathsf{H}$ in eq.\ (\ref{HK_matrix}), using a numerical
implementation of a general algorithm to diagonalize a quadratic
form of boson operators proposed in \cite{S_numeric}. Changing the
relative strength of the Kitaev term, for example $\alpha=0.4$
compared to 0.5, does not change the spectrum qualitatively only
introduces a weak dispersion in the gapped $\omega_{1,2}$ modes,
compare Figs.\ \ref{fig_disp}d-e).

The dispersions show many distinct features compared to the case
when the same stripy ground state was stabilized instead by
isotropic Heisenberg exchanges shown in Fig.\ \ref{fig_disp}g).
Notably there is no longer a gapless mode at the $\Gamma$ point
and at the Bragg peak positions (X and Y). The lowest mode softens
at the M point as in previous cases due to the 3-domain structure
of the stripy ground state. The dispersion is gapless at this
point in the linear spin-wave approximation and a gap is predicted
to open up when quantum fluctuations to 1st order in $1/S$ are
included for any general $\alpha$, except for the exactly solvable
point $\alpha=0.5$ where due to an exact cancellation the spectrum
is gapless \cite{S_jackeli}.

A spherical average of the spin-wave spectrum in Fig.\
\ref{fig_disp}d) (including both the $S^{xx}$ and $S^{yy}$
dynamical structure factors) is shown in Fig.\ 3j), the lower
boundary of the scattering at low-$Q$ (emphasized by the red solid
line) is due to scattering off magnons on the $\omega_4$
$\Gamma$-M dispersion branch near the M point.\\

{\bf S8. Spin-wave dispersions for the
Kitaev-Heisenberg-$J_2$-$J_3$ model in the zig-zag phase}\\

Here we explore the effects of adding a small Kitaev interaction
$J_{\rm K}$ to the $J_{1,2,3}$ Hamiltonian when the ground state
order is the zig-zag phase (this has recently been shown to be
stable for a range of $J_{\rm K}$ values \cite{S_Kimchi}). We
obtain the spin-wave Hamiltonian matrix in this case by combing
eqs. (\ref{Ham_matrix}) and (\ref{HK_matrix}) as
\begin{equation} \label{HKJ23_matrix}
\mathsf{H}=\frac{1}{2}\left[\begin{array}{llllllll}
 A&   B&  0&   C&  0&   D&  E&    F  \\  
 B^*& A&  C^*& 0&  D^*&   0&  F^*&  E  \\  
 0&   C&  A&   B&  E&   F&  0&    D  \\  
 C^*& 0&  B^*& A&  F^*& E&  D*&    0  \\  
 0&   D&  E&   F&  A&   B&  0&    C  \\  
 D^*&   0&  F^*& E&  B^*& A&  C^*&  0  \\  
 E&   F&  0&   D&  0&   C&  A&    B  \\  
 F^*& E&  D^*&   0&  C^*& 0&  B^*&  A  \\  
\end{array}\right]  \tag{S7}
\end{equation}
where
\begin{displaymath}
\begin{array}{l l}
    A&=S\ \{-J_1+2J_2+3J_3+2J_2\cos(2\pi h)+J_{\rm{K}}\} \\
    B&=(-1/2)SJ_{\rm{K}}{\eta}^{-2} \\
    C&=S\ \{2J_1\cos(\pi h) -(1/2)J_{\rm{K}}{\zeta} \}{\eta} \\
    D&=S\ \{J_1{\eta}^{-2}+J_3\left[{{\eta}}^{4} +2{{\eta}}^{-2}\cos(2\pi h)\right]-J_{\rm{K}}{\eta}^{-2}/2 \}\\
    E&=2SJ_2\{\cos\left[\pi(h+k)\right]+\cos\left[\pi(h-k)\right]\} \\
    F&=(1/2)SJ_{\rm{K}}{\zeta}{\eta} \\
    \zeta&=e^{h \pi i}\\
    \eta&=e^{k \pi i/3}.
\end{array}
\end{displaymath}
Diagonalization leads to four dispersion branches
\begin{equation}
\begin{array}{ll}
\omega_{1,2}^2({\bm q})&=A^2-E^2+|B-C|^2-|D-F|^2\\
                       &\pm\sqrt{4|A(B-C)+E(D-F)|^2-\delta_{-}^{2}}\\
\omega_{3,4}^2({\bm q})&=A^2-E^2+|B+C|^2-|D+F|^2\\
                       &\pm\sqrt{4|A(B+C)-E(D+F)|^2-\delta_{+}^{2}}\\
                       \end{array} \tag{S8} \label{dispHK23}
\end{equation} \\
where $\delta_{\pm}=|(B \pm C)(D^* \pm F^*)-(B^* \pm C^*)(D \pm F)|$.
The dispersions are plotted in Figs.~\ref{fig_disp}i-j) for $\alpha$=0.4 ($J_{\rm
K}/J_1=4/3$), $J_2$/$J_1$~=0.23 and $J_3$/$J_1$~=0.51. To discuss
the key features of the spectrum it is helpful to visualize the
degeneracies associated with the magnetic order. The magnetic
structure is the zig-zag pattern shown in Fig.\ \ref{fig_disp}a)
but where the spin direction can be either along the $\hat{\bm x}$
direction to satisfy the Kitaev term on the $x$-type AD bond, or
along the $\hat{\bm y}$ direction to satisfy the Kitaev exchange
on the $y$-type BC bond. At the classical level any in-between
direction, i.e. in the  $\hat{\bm x}\hat{\bm y}$ plane, also has
the same energy, so one expects a gapless mode associated with
rotations in this ``easy" plane. Indeed Fig.\ \ref{fig_disp}j)
shows that the dispersion is gapless at the Y point and with
strong intensity for fluctuations in this easy-plane (along the
$\hat{\bm y'}$ normal to the ordered direction labelled $\hat{\bm
x'}$), and gapped for fluctuations along $\hat{\bm z}$ out of the
easy plane, see Fig.\ \ref{fig_disp}i). Furthermore, due to the
honeycomb lattice geometry the magnetic structure is degenerate
with another two domains rotated by $\pm60^{\circ}$ around the
axis normal to the plane, so the spectrum is gapless at the Bragg
peak positions of those other two domains, at points equivalent to
M. The Hamiltonian however does not poses any continuous
rotational symmetry in the presence of the Kitaev term, so one
might expect that small gaps would open at both Y and M points
when quantum fluctuations are included so the spectrum would be
fully gapped. For completeness we quote the Curie-Weiss
temperature for this model $\theta_{\rm
CW}=-S(S+1)(J_1+2J_2+J_3-J_{\rm K}/3)/k_{\rm
B}$.\\

{\bf S9. Spherically-averaged neutron scattering intensity}\\

The one-magnon neutron scattering intensity including the magnetic
form factor and neutron polarization factor is proportional to
\begin{equation}
\left(\frac{g}{2} f(Q)\right)^2\left[
 \left( 1-\frac{Q_x^2}{Q^2}\right) S^{xx}(\bm Q,\omega)
+\left( 1-\frac{Q_y^2}{Q^2}\right) S^{yy}(\bm Q,\omega)\right]
\nonumber \tag{S9} \label{intensity}
\end{equation}
where we used for $f(Q)$ the Ir$^{4+}$ spherical magnetic form
factor \cite{S_Irmff} and assumed the $g$-factor equal to 2. Here
$Q_x$ ($Q_y$) are the components of the wavevector transfer ${\bm
Q}$ along the $x$-axis ($y$-axis), where $z$ is the ordered spin
direction. The precise direction of the ordered moments ($\hat{\bm
z}$-axis) with respect to the crystallographic axes has only a
small effect on the powder-averaged spectrum via small intensity
modulations through the polarization factors $\left(
1-\frac{Q_{x,y}^2}{Q^2}\right)$, however for concreteness, we
included a specific moment direction for the comparison with data.
For the $J_{1,2,3}$ model in Figs.\ 3(h-i) the moments were
assumed to be aligned along the crystallographic ${\bm a}$-axis
(as suggested by resonant xray data \cite{S_hill}) and for the KH
model [Fig.\ 3j)] the moment is assumed to be along the cubic
$\hat{\bm z}$-axis closest to the ${\bm a}$-axis (tilted
out-of-plane by 35.26$^{\circ}$ from the ${-\bm a}$ axis, see
Fig.\ \ref{fig_disp}b) inset). Eq.\ (\ref{intensity}) was
numerically averaged over a spherical distribution of orientations
for the wavevector transfer $\bm{Q}$ and convolved with the
instrumental resolution to obtain the plots in Figs.\ 3h-j),
directly comparable with the raw neutron scattering data in Fig.\
3e). For the KH-$J_2$-$J_3$ model the intensity is also given by
eq. (\ref{intensity}) but with the axis labels ($x$,$y$,$z$)
replaced by ($y'$, $z$, $x'$), where the $x'$-axis defines the
ordered moment direction (located in the original $xy$ plane) and
$y'$ and $z$ are orthogonal directions to it.


\end{document}